\documentclass[aip,reprint]{revtex4-1}

\usepackage{amsmath}
\usepackage{graphicx}
\usepackage[textwidth=180mm]{geometry}
\usepackage{dcolumn}
\usepackage{bm}
\usepackage{verbatim}

\usepackage[valuesep=thin, unitsep=medium, digitsep=thin, per-mode=symbol]{siunitx}

\usepackage[T1]{fontenc}
\DeclareSIUnit[number-unit-product = {}]{\inchQ}{\textquotedbl}
\DeclareSIUnit[number-unit-product = {\thinspace}]{\inch}{in}

\usepackage{xcolor}
\definecolor{desyorange}{RGB}{242,142,0}
\definecolor{desyblue}{RGB}{0,166,235}
\usepackage[normalem]{ulem}

\begin{document}

\title{The Phase-Contrast Imaging Instrument at the Matter in Extreme Conditions Endstation at LCLS}

\keywords{LCLS, MEC, x-ray FEL, Phase Contrast Imaging, HEDS}
\pacs{07.85.Qe,07.85.Tt,61.05.cp, 62.50.Ef , 62.50.-p, 64.30.Ef,}

\author{Bob Nagler}
\email{BNagler@slac.stanford.edu}
\affiliation{SLAC National Accelerator Laboratory, 2575 Sand Hill
  Road, Menlo Park, CA 94025, USA} 
\author{Andreas Schropp}
\affiliation{Deutsches Elektronen-Synchrotron (DESY), Notkestr. 85, D-22607 Hamburg, Germany}
\author{Eric C. Galtier}
\affiliation{SLAC National Accelerator Laboratory, 2575 Sand Hill Road, Menlo Park, CA 94025, USA} 
\author{Brice Arnold}
\affiliation{SLAC National Accelerator Laboratory, 2575 Sand Hill Road, Menlo Park, CA 94025, USA} 
\author{Shaughnessy B. Brown}
\affiliation{Stanford University, 450 Serra Mall, Stanford, CA 94305, USA}
\affiliation{SLAC National Accelerator Laboratory, 2575 Sand Hill Road, Menlo Park, CA 94025, USA} 
\author{Alan Fry}
\affiliation{SLAC National Accelerator Laboratory, 2575 Sand Hill Road, Menlo Park, CA 94025, USA} 
\author{Arianna Gleason}
\affiliation{Los Alamos National Laboratory, Los Alamos, NM 87545, USA}
\affiliation{SLAC National Accelerator Laboratory, 2575 Sand Hill Road, Menlo Park, CA 94025, USA} 
\author{Eduardo Granados}
\author{Akel Hashim}
\author{Jerome B. Hastings}
\affiliation{SLAC National Accelerator Laboratory, 2575 Sand Hill Road, Menlo Park, CA 94025, USA} 
\author{Dirk Samberg}
\author{Frank Seiboth}
\affiliation{Institute of Structural Physics, Technische Universit\"at Dresden, 01062 Dresden, Germany}
\author{Franz Tavella}
\author{Zhou Xing}
\affiliation{SLAC National Accelerator Laboratory, 2575 Sand Hill Road, Menlo Park, CA 94025, USA} 
\author{Hae Ja Lee}
\affiliation{SLAC National Accelerator Laboratory, 2575 Sand Hill Road, Menlo Park, CA 94025, USA} 
\author{Christian G. Schroer}
\affiliation{Deutsches Elektronen-Synchrotron (DESY), Notkestrasse 85, D-22607
  Hamburg, Germany}
\affiliation{Department Physik, Universit\"at Hamburg, Luruper Chaussee 149, D-22761 Hamburg, Germany.}

 \begin{abstract}
We describe the Phase-Contrast Imaging instrument at the Matter in Extreme Conditions (MEC) endstation of the Linac Coherent Light Source. 
The instrument can image phenomena  with a spatial resolution of a few hundreds of nanometers and at the same time reveal the atomic structure through X-ray diffraction, with a temporal resolution better than \SI{100}{\femto\second}.
It was specifically designed for studies relevant to High-Energy-Density Science and can monitor, e.~g., shock fronts, phase transitions, or void collapses. This versatile instrument was commissioned last year and is now available to the MEC user community.
\footnote{published in Rev. Sci. Instrum. \textbf{87}, 103701 (2016) : http://dx.doi.org/10.1063/1.4963906}
\end{abstract}

\date{\today}

\maketitle

\section{Introduction \label{sec:intro}}

Hard x-ray radiography has been widely used in the past to investigate High-Energy-Density phenomena such as radiatively driven shock waves~\cite{Hammel1993}, Inertial Confinement Fusion capsule implosions~\cite{Katayama1993,Kalantar1997,Marshall2009,Hicks2010}, X-pinch plasmas~\cite{Shelkovenko2001} and other hydrodynamical evolution of targets under test (see Landen, \textit{et al.} and references therein~\cite{Landen2001}). The material transformation of such targets tends to be very rapid. Therefore, short x-ray pulses are needed, which have historically been produced by high power lasers, with spatial and temporal resolution of the order of tens of micrometers and hundreds of picoseconds or more, respectively.

At the same time, techniques that exploit the phase changes of x rays while they propagate through matter were developed~\cite{Davies:1984p1835}. At x-ray photon energies, the real part of the index of refraction of matter is typically much larger than its imaginary part. Therefore, Phase-Contrast Imaging (PCI) techniques are often considerably more sensitive than methods that rely on absorption alone, such as x-ray radiography. Such PCI techniques have also been adapted for use with laser-based backlighters~\cite{Montgomery2004,Kozioziemski2005,Koch2009,Ping2011}. However, due to their spatial incoherence and low spectral brightness, the spatial resolution remained limited. In-line geometries that do not require x-ray optics but rely on free propagation of the electromagnetic waves were conceived, and sub-micron resolution are now standard at synchrotron radiation facilities~\cite{Snigirev1995,Nugent1996,Davis1995,Cloetens1996,Cloetens1999}.

The advent of x-ray free-electron lasers (XFEL) in general and the Linac Coherent Light Source (LCLS)~\cite{Emma2010,Pellegrini2016,Bostedt2016} in particular has opened new possibilities for phase-contrast imaging in High-Energy-Density Science (HEDS). The spatial coherence of the LCLS beam and its focusability allows for imaging with sub-micron spatial resolutions as in synchrotron facilities, while the short, bright pulses allow temporal resolution of tens of femtoseconds, faster than any phonon timescale. Furthermore, the Matter in Extreme Conditions (MEC) endstation~\cite{Nagler2015} is specifically tailored to field experiments in HEDS. 

However, an easy transfer of phase-contrast imaging methods established at synchrotron radiation sources to the XFEL is not possible as they often rely on the measurement of a series of images for different propagation distances between sample and detector~\cite{Cloetens1999}. In most HED pump-probe experiments the sample is destroyed by the intense optical pump pulse, which prevents to repeat an imaging experiment on the same sample. For this reason, in first PCI-experiments at the Linac Coherent Light Source (LCLS) an iterative phase retrieval scheme was followed, which often allows to retrieve a quantitative measure of the phase shift introduced by an object from a single phase-contrast image~\cite{Giewekemeyer2011}. The method is especially successful if strong constraints on the object's transmission function can be applied during the iterative phase retrieval, such as, e.~g., in the case of pure phase objects with negligible absorption. In this way, experiments  that image shock waves in diamond using phase contrast with unprecedented spatial and temporal resolution have been performed\cite{Schropp2015}. The spatial resolution of PCI is currently mainly limited by the SASE bandwidth of the LCLS-pulse smearing out features smaller than a few hundred nanometers. In addition, PCI can easily be combined with X-ray diffraction. In this way, macroscopic imaging with a resolution of hundreds of nanometer and information about the atomic structure are recorded simultaneously, with a time resolution better than \SI{100}{\femto\second}.

To foster this field of research, a PCI instrument was designed for MEC, and is now available to the wider user community. In this paper, we describe this new PCI instrument. In section~\ref{sec:description}, we give a brief overview of the MEC endstation, the  Beryllium Compound Refractive Lenses (Be CRLs) that are used to focus the LCLS beam, and both the ptychographic and phase-contrast imaging methods we use. In section~\ref{sec:com} we show some commissioning results that illustrate the capabilities of the PCI instrument, and present an outlook in section~\ref{sec:conclusions}.

\section{Instrument description \label{sec:description}}
\subsection{A short overview of the MEC beamline\label{sec:mec}}

The Matter in Extreme Conditions endstation is one of seven endstations at the Linac Coherent Light Source\cite{Emma2010,Pellegrini2016}. It is equipped with high-power lasers and diagnostics to study high-pressure science, shock-induced chemical reactions and phase changes, dislocation dynamics, high strain rate phenomena, material strength, warm dense matter and dense plasmas\cite{Nagler2015,Bostedt2016}. To this end, two optical laser systems are available at MEC: a nanosecond glass laser, generally used as a shock driver, that provides two beams of \SI{20}{\joule} each with a pulse length varying from \SI{2}{\nano\second} to \SI{200}{\nano\second}. The laser can be operated at \SI{0.002}{\hertz} and the temporal profile of the pulse can be shaped in the laser front-end. A short pulse Ti:Sapphire system (\SI{45}{\femto\second}) is present that can deliver \SI{1}{\joule} at \SI{5}{\hertz} or \SI{4}{\joule} at \SI{0.002}{\hertz}.

\begin{figure}[bt]
  \centering
  \includegraphics[width=9cm]{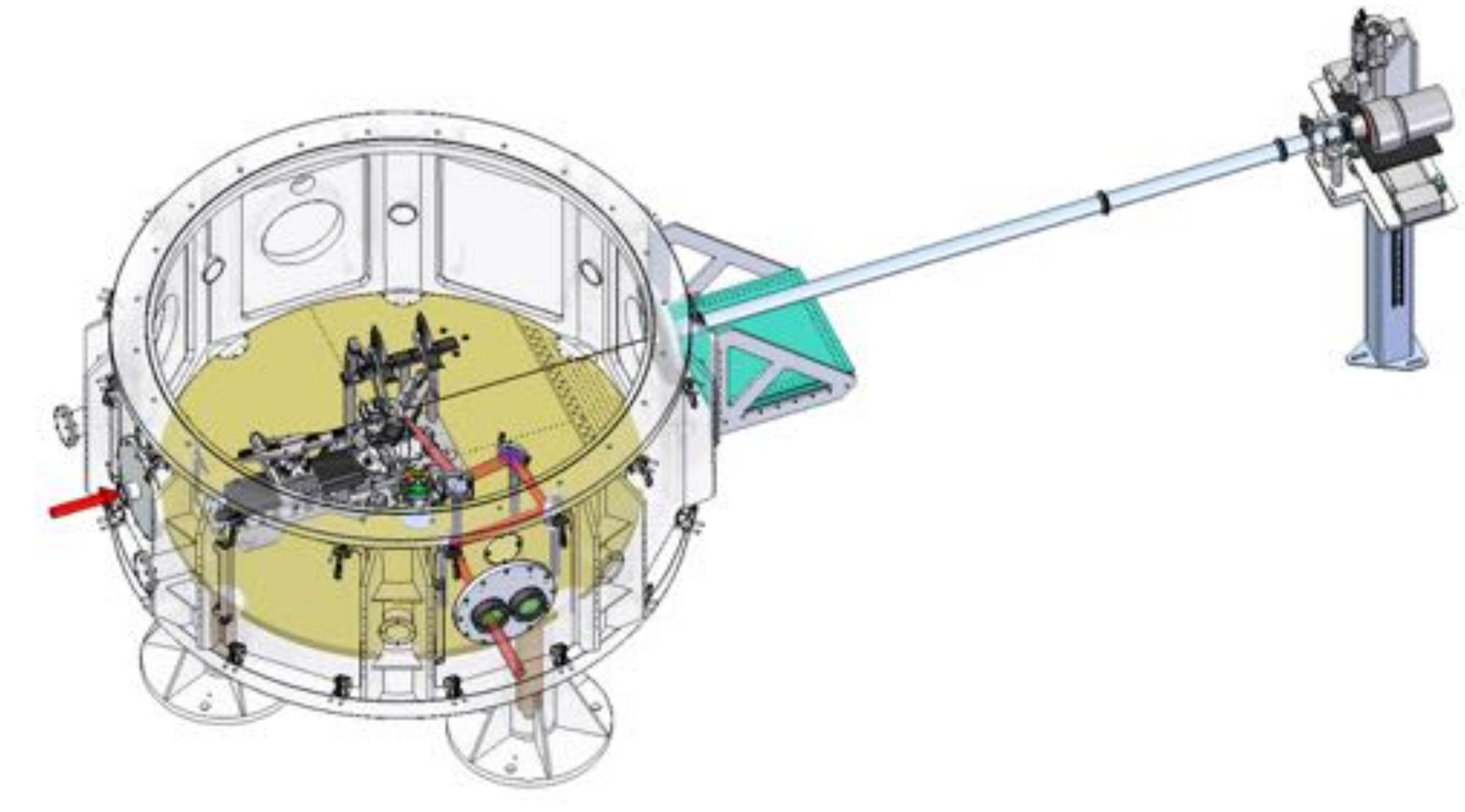} 
  \caption{3D model of the PCI instrument in the MEC target chamber, and location of the x-ray cameras behind the MEC chamber. Different x-ray cameras can be translated into the beam on a motorized stage. Camera location can be varied between \SI{1.2}{\meter} and \SI{5}{\meter} from the target, by changing the length of the flight tube.}
  \label{fig:overview}
\end{figure}

\begin{figure}[bt]
  \centering
  \includegraphics[width=87mm]{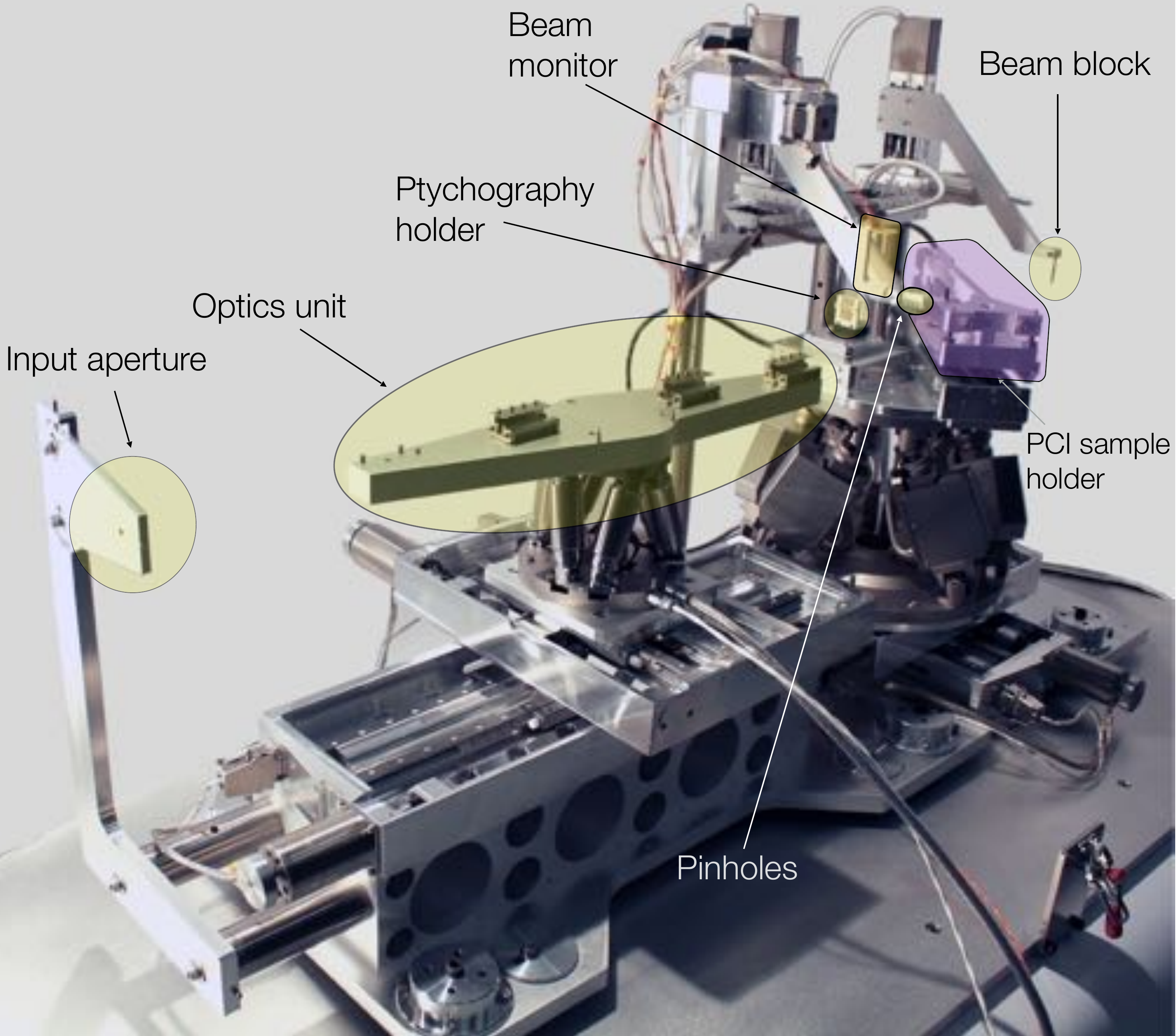}
  \caption{Picture of the  Phase-Contrast Imaging instrument at MEC. }
  \label{fig:pciinstrument}
\end{figure}

The MEC target chamber is located approximately \SI{460}{\meter} from the end of the LCLS undulator. A SiC coated hard x-ray offset  mirror system (HOMS), positioned at a distance of \SI{156.07}{\meter} from the source, steers approximately \si{10^{12}} hard x-ray photons, tunable in energy from \SI{4}{\kilo\electronvolt} to \SI{12}{\kilo\electronvolt} in the fundamental, to the endstation. The LCLS beam has an intrinsic relative bandwidth of \si{10^{-3}}. While this is sufficient for the Phase Contrast Imaging experiment, it does pose a problem for the ptychographic spot size characterization described below, in which case a monochromator that reduces the bandwidth to $1 \times 10^{-4}$ is used. The X-ray pulse length is typically 60~fs (but can be tuned to smaller than 10~fs), leading to high temporal resolution. Multiple diagnostics (e.~g., profile monitors, energy monitors, timing systems) are placed at various locations along the beamline. The nominal beam size at MEC is around \SI{1}{\milli\meter}, and can be apertured by slits. The standard focusing system in use at MEC consists of Be CRLs that are located \SI{4}{\meter} upstream of the Target Chamber Center (TCC). 
The MEC target chamber is a cylindrical vacuum vessel with diameter of approximately \SI{2}{\meter}. It is separated from the rest of the beamline by a \SI{20}{\micro\meter} thick beryllium window. After interaction with the sample at TCC, the x-rays  go through a flight tube at the back of the chamber. A \SI{100}{\micro\meter} Kapton\textsuperscript{{\textregistered}} window isolates the vacuum at the end of the flight tube. X-ray cameras can be placed on a \SI{500}{\milli\meter} travel stage behind the window, and positioned into the beam to record images of the x-ray beam or small angle x-ray scattering. The distance from TCC to the x-ray camera can range from \SI{1.2}{\meter} to \SI{5}{\meter}, although it is in principle possible to put a vacuum compatible camera in the MEC chamber and forego the flight tube. A general overview of the location of the MEC target chamber with the PCI instrument and x-ray cameras can be seen in Fig.~\ref{fig:overview}, while a close-up of the instrument itself is shown in  Fig.~\ref{fig:pciinstrument}.

A more detailed description of the MEC endstation can be found in Nagler, \textit{et al.}\cite{Nagler2015}.

\subsection{The Phase-Contrast Imaging Instrument at MEC}

The PCI instrument can be installed in the vacuum vessel of MEC. A picture can be seen in Fig.~\ref{fig:pciinstrument}). 
The instrument is designed to operate in two different modes:
\begin{itemize}
	\item magnified-phase contrast imaging of targets with XFEL pulses in optical-pump-x-ray-probe experiments.
	\item ptychographic imaging of test objects in order to accurately determine the spatial profile of the focused XFEL beam. The precise knowledge of the wave field is needed for a quantitative analysis of PCI data.
\end{itemize}

\subsubsection*{Optics unit}

The optics unit is designed to hold a maximum of four lens stacks and can be aligned with six degrees of freedom. A longer linear stage (PI Micos HPS-170 \SI{12}{\inchQ}, \SI{300}{\milli\meter} travel range) allows to move the lenses along the LCLS beam and a smaller stage (PI Micos, HPS-170 \SI{4}{\inchQ}, \SI{100}{\milli\meter} travel range) is used to move the whole platform perpendicular to the LCLS beam in order to switch between the different lens sets. The hexapod (PI M-811.DV2) on top these stages provides the missing degrees of freedom to accurately position the lenses (linear motion in $z$-direction and three rotations). Travel ranges of this device are specified as $\pm \SI{17}{\milli\meter}$, $\pm \SI{16}{\milli\meter}$, $\pm \SI{6.5}{\milli\meter}$, $\pm \SI{10}{\degree}$, $\pm \SI{10}{\degree}$, $\pm \SI{21}{\degree}$ in $x$, $y$, $z$, $\vartheta_z$, $\vartheta_y$, and $\vartheta_z$, respectively.

\subsubsection*{Sample/target unit}

The sample/target stage is designed to accommodate simultaneously the targets for optical-pump-x-ray-probe experiments and a set of nanostructured test objects on a high-precision scanning stage for ptychographic beam characterization. Switching between these two operation modes can be done within minutes, so the nanobeam can be characterized in detail directly prior to PCI experiments.

The samples are placed on top of 2 linear stages (PI Micos HPS-170, \SI{100}{\milli\meter} travel range), which are used to switch between the ptychography configuration (cf.~Sec.~\ref{sec:ptycho}) and the PCI configuration (cf.~Sec.~\ref{sec:pci}). A hexapod (PI Micos, H-824 GV2), with 6 degrees of freedom and sub-micron resolution and repeatability, is used for precise alignment of the sample. Travel ranges are $\pm \SI{22.5}{\milli\meter}$, $\pm \SI{22.5}{\milli\meter}$, $\pm \SI{12.5}{\milli\meter}$, $\pm \SI{7.5}{\degree}$, $\pm \SI{7.5}{\degree}$, $\pm \SI{12.5}{\degree}$ in $x$, $y$, $z$, $\vartheta_z$, $\vartheta_y$, and $\vartheta_z$, respectively. On top of the hexapod a piezo stage (PI Micos P-733.3VD, closed loop travel ranges are \SI{100}{\micro\meter} in $x$ and $y$, as well as \SI{10}{\micro\meter} in $z$ direction, respectively) is implemented to scan ptychography samples with nanometer precision (repeatability $< \SI{2}{\nano\meter}$ for all axes). In PCI mode another linear stage (PI Micos PLS-85, \SI{155}{\milli\meter} travel range) is mounted at \SI{45}{\degree} relative to the LCLS beam,  to easily switch between sample cassettes (cf.~Fig.~\ref{fig:pci1}).

\subsubsection*{Pinhole and beam stop}

A pinhole can be inserted directly in front of a sample in order to clean the focused beam from background radiation. It is held by a set of three linear stages, one Aerotech MPS50SL-050-VAC7-SM stage  with a travel range of \SI{50}{\milli\meter} for movements in $x$-direction and two Aerotech MPS50SL-025-VAC7-SM stages with a travel range of \SI{25}{\milli\meter} to position the pinhole in $y$- and $z$-direction, respectively. The beam stop positioning after the sample is implemented using the same configuration of stages as for the pinhole.

\subsubsection*{Beam monitor}
A beam monitor  (see Fig.~\ref{fig:pciinstrument}) is placed after the Be lenses and before the pinholes and PCI or ptychography sample. Its working principle is similar to the intensity and position monitors (IPMs) that are used in the hard x-ray beamlines, but without the position sensitivity\cite{Feng2011,Tono2011}. The x-ray beam passes through a thin foil that is positioned in the x-ray beam. Four foils of different thickness and composition can be chosen. The foil scatters a small fraction of the beam or creates K-$\alpha$ x-ray radiation. These x-rays are recorded with an x-ray diode (Canberra C14560-2 RF14*14-300EB) every shot, resulting in a signal that is proportional to the incoming x-ray beam, and can be used in data analysis.

\subsection{Creating a secondary source for magnified phase-contrast imaging using Be CRLs\label{sec:crl}}

\begin{figure}[bt]
  \centering
  \includegraphics[width=9cm]{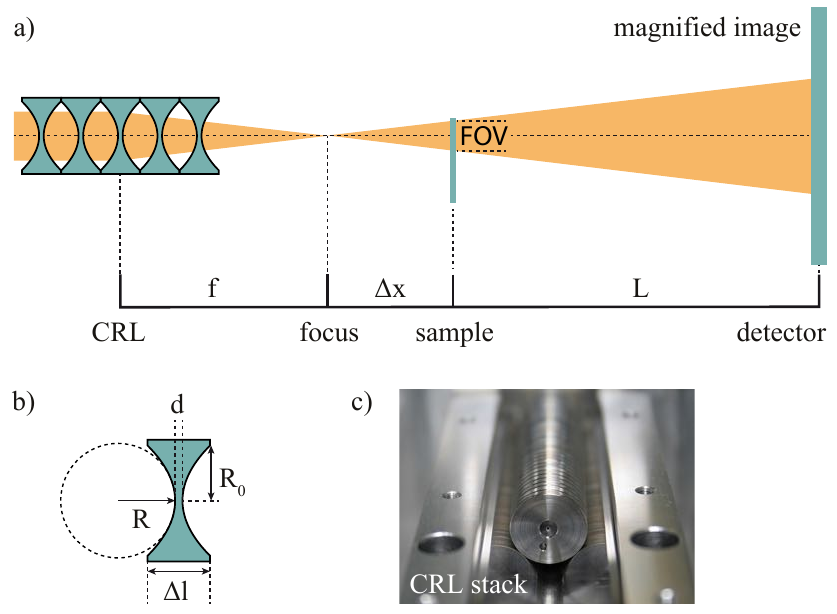} 
  \caption{a) Schematic of the Be CRL setup in PCI geometry. The lens stack has a focal length $f$. The focus is placed at a distance $\Delta x$ in front of the sample. A detector is then positioned a distance $L$ after the sample, leading to a magnification $M = (L+\Delta x)/{\Delta x}$. b) Sketch of a single parabolic Be CRL indicating relevant geometric parameters. c) Image of a stack of Be CRLs aligned in a lens holder.}
  \label{fig:belenses}
\end{figure}

The PCI instrument uses parabolic compound refractive lenses (CRLs) from beryllium to focus the LCLS beam. The fabrication and  working mechanism of CRLs, and their uses in x-ray microscopy  have been described extensively in literature\cite{Snigirev1996,Lengeler1999a,Lengeler1999b,Lengeler2002}. In short, beryllium lenses with a concave parabolic profile are stacked (see Fig.~\ref{fig:belenses}). The focal length, $f$, of the stack is equal to 

\begin{equation}
f\simeq\frac{R}{2N\delta}, 
\end{equation}

with $n=1-\delta+i\beta$ the refractive index of the lenses, $N$ the number of lenses, and  $R$ the radius of curvature at the vertex of an individual lens. For magnified phase-contrast imaging, we place a sample at distance $\Delta x$ behind the focus and the x-ray detector a distance $L$ behind the sample. This gives a magnification $M=(L+\Delta x)/\Delta x$ and a field of view in the sample plane of $FOV = D_{\rm eff}/f\cdot \Delta x$, with $D_{\rm eff}$ being the effective aperture of the lens stack \cite{Lengeler1999b}. In order to obtain a large magnification and field of view, a large numerical aperture and thus a small focal length of the optics is needed. Therefore, beryllium lenses with a radius $R=\SI{50}{\micro\meter}$ and geometric aperture $D = 2R_0 =\SI{300}{\micro\meter}$ are typically chosen. Here, $D$ is the diameter and $R_0$ the radius of the geometric aperture. Stacking between 15 and 30 of such lenses, focal lengths as small as \SI{200}{\milli\meter} for photon energies ranging from \SI{4}{\kilo\electronvolt} to \SI{9}{\kilo\electronvolt} can be achieved, generating a focus with a lateral size in the range of \SI{100}{\nano\meter}.
 
The x-ray focus serves as a secondary source for magnified imaging (cf.~Fig.~\ref{fig:belenses}). While the distance $\Delta x$ can be freely chosen, $L$ is limited at MEC to about \SI{5}{\meter}, due to the size of the MEC hutch. 
In the PCI instrument the Be CRLs are kinematically mounted on top of the optics unit described above.
 The travel range of the stage  along the x-ray beamline allows us to change the magnification and field of view online by effectively changing the distance $\Delta x$. 
The four CRL stacks that can be mounted on the optics table allow distances between the lens and the target ranging from \SI{100}{\milli\meter} to \SI{800}{\milli\meter}. In this way, a large field of view of up to \SI{1.2}{\milli\meter} can be reached for a lens stack with \SI{200}{\milli\meter} focal length.

\subsection{Ptychography\label{sec:ptycho}}

\begin{figure}[bt]
\centering
\includegraphics[width=9cm]{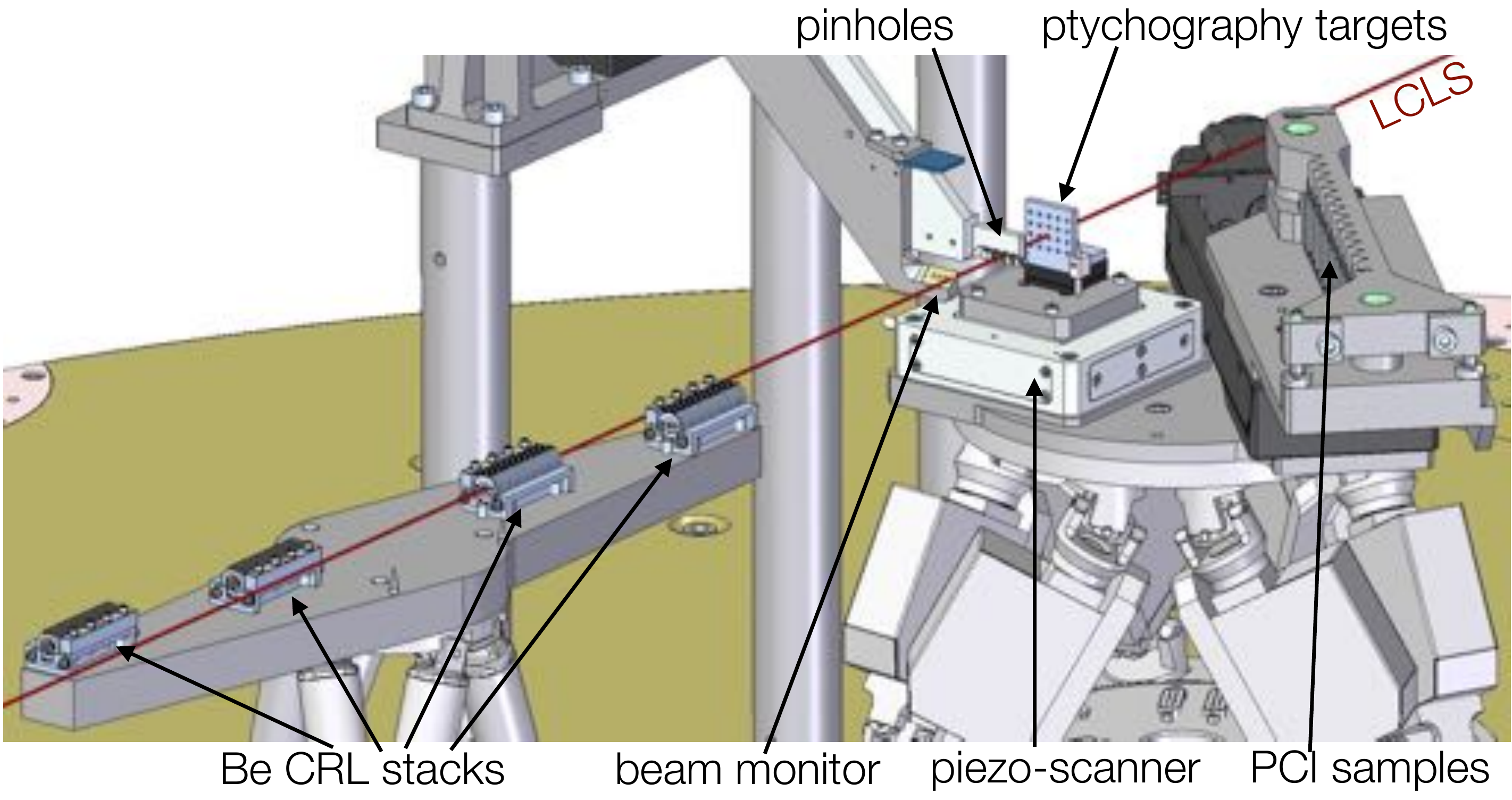}
\caption{Ptychographic sample environment. A sample with nanometer
  sized structures is positioned close to the x-ray focus and normal to
the beam. It is located on a 3-axis piezo-drive stage (PI P-733) that can scan
the sample through the beam with a nm repeatability and sub-nanometer
resolution. Pinholes of varying sizes can be
placed in front of the target. }
\label{fig:ptycho1}
\end{figure}

In order to interpret the phase-contrast images quantitatively, it is crucial to know the illuminating x-ray field incident on a sample in both amplitude and phase. The field can be determined with the PCI instrument using scanning coherent x-ray diffraction also known as ptychography~\cite{Rodenburg2004,Thibault2008}. A nano-structured object is scanned through the x-ray beam, and a diffraction pattern is recorded at every scan point. Using  phase retrieval algorithms the complex-valued transmission function of the nanostructured object as well as the illuminating x-ray field can be reconstructed~\cite{MR2009}. Ptychography is routinely used at synchrotron facilities to characterize focusing optics and nanobeams~\cite{Schropp2010,Kewish2010,Hoenig2011} and has been successfully applied at LCLS\cite{Schropp2013}. Fig.~\ref{fig:ptycho1} shows the instrument in ptychographic imaging mode. To acquire a ptychographic data set, the sample is scanned through the beam with a piezo driven 3-axis stage (PI P-733), while far-field diffraction images are captured on an x-ray detector (e.~g., a Cornell-SLAC hybrid Pixel Array Detector (CSPAD) or ePix camera\cite{Blaj2015}) located outside the chamber, typically at a distance of \SI{4.8}{\meter} (not shown). Pinholes on a 3D alignment stage can be placed before the sample to reduce diffuse scattering originating from the Be CRL or the beryllium window in the beamline. An aperture with a diameter of \SI{1.5}{\milli\meter} placed before the beryllium lenses blocks wider angle scattering from the beryllium window.

\begin{widetext}
  \begin{figure*}
    \centering
    \includegraphics[width=18cm]{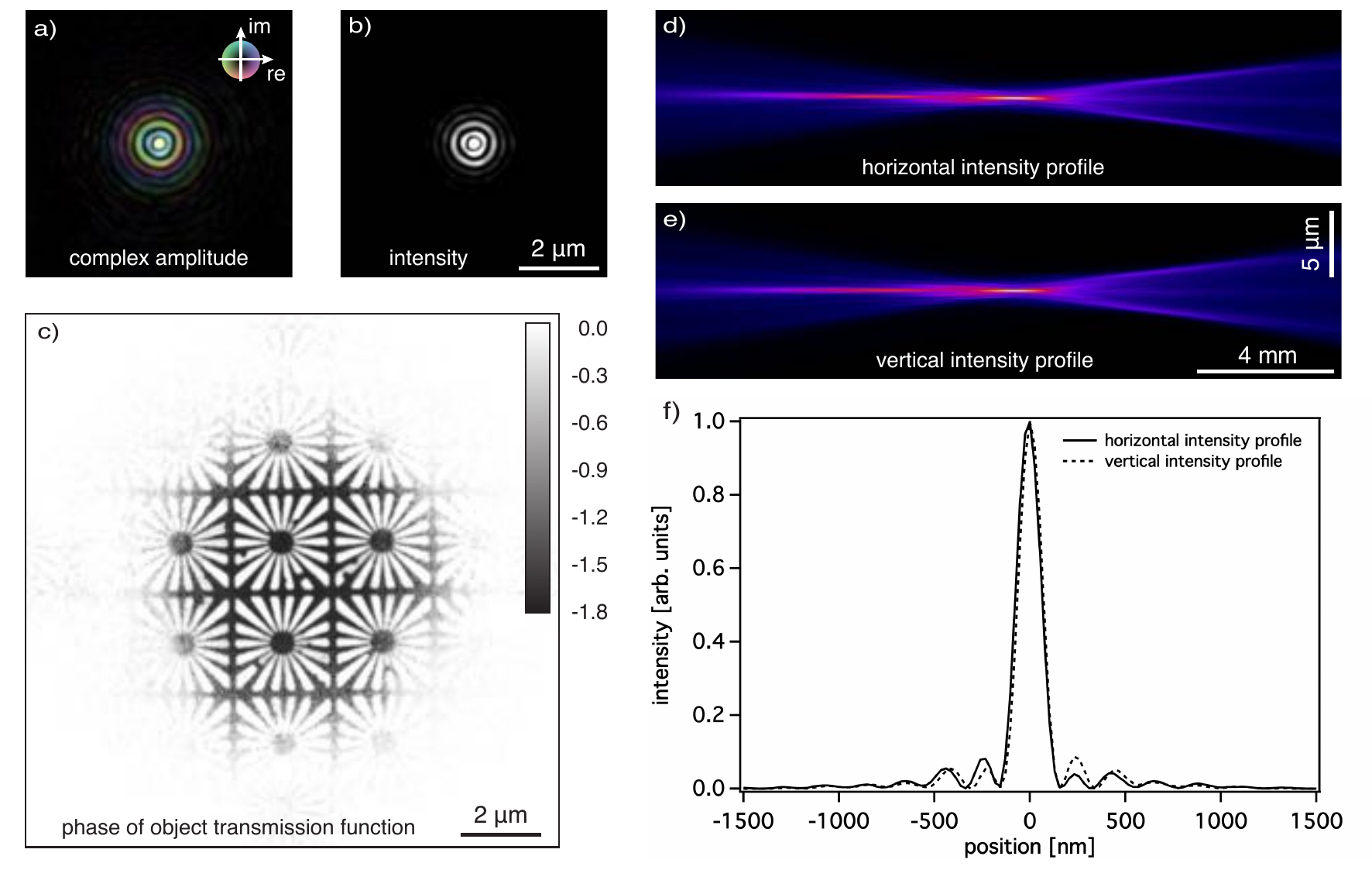}
    \caption{\label{fig:ptychorecon} Ptychographic reconstruction of a nano-structured object. a) Reconstructed complex-valued illumination function in the object plane and b) corresponding intensity distribution. c) Reconstructed ptychography sample. d) Horizontal cross-section of the x-ray intensity vs. propagation distance. e) Vertical cross section of the x-ray intensity vs. propagation distance.  f) Horizontal and vertical intensity profiles in the focal plane. The reconstructed focus has a spot size of \SI{150}{\nano\meter} (FWHM). 25 Be lenses with $R=\SI{50}{\micro\meter}$ were used and a photon energy of $E = \SI{8.2}{\kilo\electronvolt}$.}
  \end{figure*}
\end{widetext}

The nano-structured samples used for ptychography contained a matrix of $\si{10} \times \si{10}$ similar Siemens stars with a size of about \SI{2}{\micro\meter} each and feature sizes between \SI{50}{\nano\meter} and \SI{200}{\nano\meter}. These structures were etched into a \SI{1}{\micro\meter} thick tungsten layer, deposited on a \SI{100}{\micro\meter} thick CVD diamond substrate. The sample was scanned through the coherent focused LCLS beam, while collecting far-field diffraction patterns with a CSPAD at each scan point. In the presented measurement (see~Fig.~\ref{fig:ptychorecon}) the sample was moved continuously by about \SI{2}{\micro\meter} in the vertical direction with a speed of $v_s = \SI{1.2}{\micro\meter\per\second}$ while LCLS was running at a repetitation rate of \SI{120}{\hertz}. One scan line in vertical direction contained 203 scan points with a distance of about \SI{10}{\nano\meter} between them. The same scan was repeated 20 times with an \SI{100}{\nano\meter} offset in horizontal direction between vertical scan lines. In this way, a total of 4060 diffraction patterns were recorded over a scan area of $\SI{2}{\micro\meter} \times \SI{2}{\micro\meter}$ in less than a minute. These diffraction patterns were sorted by their integral intensity and the \SI{20}{\percent} with highest as well as the \SI{60}{\percent} with lowest intensity were rejected, i.~e., only \SI{20}{\percent} of them with an intermediate intensity were used during the subsequent phase retrieval. In ptychography mode the sample needs to be positioned close to the focal plane. Given the experimental parameters of photon energy $E=\SI{8.2}{\kilo\electronvolt}$ (wavelength $\lambda = hc/E$ = \SI{1.51}{\angstrom}) and the distance between sample and detector $L=\SI{4815}{\milli\meter}$, the real space pixel size is calculated to $p_r = \lambda L/(Np_d) = \SI{25.9}{\nano\meter}$, with $p_d=\SI{110}{\micro\meter}$ the pixel size of the CSPAD and $N = 256$ the size of the pixel subarea used for the reconstruction. A summary of a typical result is shown in Fig.~\ref{fig:ptychorecon} showing the reconstructed complex-valued illumination electromagnetic field [cf.~Fig.~\ref{fig:ptychorecon}\,a)], its intensity [cf.\,Fig.~\ref{fig:ptychorecon}\,b)], the phase of the complex-valued transmission function of the object [cf.\,Fig.~\ref{fig:ptychorecon}\,c)], the caustic of the focused beam in horizontal and vertical direction [cf.\,Figs.~\ref{fig:ptychorecon}\,d), e)] and an intensity profile of the focused LCLS-beam [cf.\,Fig.~\ref{fig:ptychorecon}\,f)] demonstrating in this case a focused x-ray beam with a size of about \SI{150}{\nano\meter} (FWHM).

Spot sizes that can be achieved have a nearly diffraction limited central peak, and sizes as small as \SI{100}{\nano\meter} have been demonstrated~\cite{Schropp2013,Nagler2015}. Chromatic aberrations of refractive lenses can affect the focus~\cite{SSHMPLNGAZHNUVHS2014}, in particular in SASE mode at LCLS with a relative bandwidth of $\Delta E/E \approx 2\times 10^{-3}$.  All ptychographic beam characterizations at LCLS have therefore been carried out with a Bartels type monochromator at $E = \SI{8.2}{\kilo\electronvolt}$. Only recently has ptychography been extended to be feasible with polychromatic x-rays\cite{EDCSPT2014}. Whether this method can be extended to characterize the SASE nanobeams generated with refractive lenses at LCLS is still an open question.

\subsection{PCI\label{sec:pci}}

\begin{figure}[bt]
\centering
\includegraphics[width=9cm]{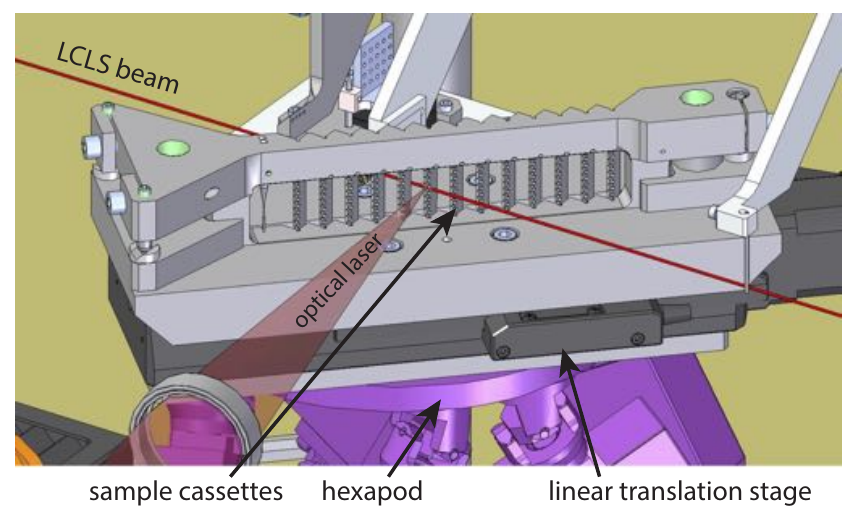} 
\caption{Picture of the PCI target mount. Targets are mounted on cassettes that load onto the target holder. The holder is designed in such a way that the optical laser can drive a shock perpendicular to the imaging direction.}
\label{fig:pci1}
\end{figure}

The targets for phase-contrast imaging are mounted on a cassette and loaded onto a saw-tooth mount that can be kinematically replaced for easy target exchange (cf.~Fig.~\ref{fig:pci1}). The mount and target holder are designed such that the x-rays image material that is driven with an optical laser either orthogonal to the x-ray direction (as in Fig.~\ref{fig:pci1}) or nearly collinear with the x-rays. The target mount is located on a linear stage (PI PLS-85) that can raster different cassettes into the beam. The hexapod, beneath the rastering stage, is used for fine alignment of the targets. The orthogonal geometry allows for imaging the propagation of shocks or hydrodynamic instabilities as they propagate in space and time. The pinholes mentioned in section~\ref{sec:ptycho} can be used to clean up the beam. A thin SiN foil, that scatters a tiny fraction of the x rays towards a diode, can be inserted right before the sample to have a relative energy measurement for every LCLS pulse.

The images are recorded on an x-ray camera which is placed outside of the target chamber, at a distance ranging between \SI{1.2}{\meter} and \SI{5}{\meter} after the flight tube mentioned in section~\ref{sec:mec}. Typically, a Ce:YAG screen coupled to an optical microscope is used to image the divergent x-ray beam [cf.~Fig.~\ref{fig:pci2}\,a)]: the compromise between magnification of the PCI setup and field of view, as shown in Fig.~\ref{fig:belenses}, limits the magnification to approximately $\times$ 20. To approach resolution of 200nm (a limit imposed by the x-ray focal spot-size), we need an imaging resolution at the camera of approximately 4 $\mu$m, which is easily achievable with an optical microscope coupled to a Ce:YAG screen. On the other hand, direct detection cameras (such as  CSPAD-140k camera\cite{Blaj2015}, PI PIXIS-XF 2048B) have a much larger resolution due to their pixel size (110 $\mu$m and 13$\mu$m respectively), but a higher sensitivity and dynamic range, and therefore can be more appropriate for highly absorptive samples if the smaller resolution (or a smaller FOV) is acceptable.

\section{Commissioning results  \label{sec:com}}

\begin{figure}[bt]
 \centering
 \includegraphics[width=9cm]{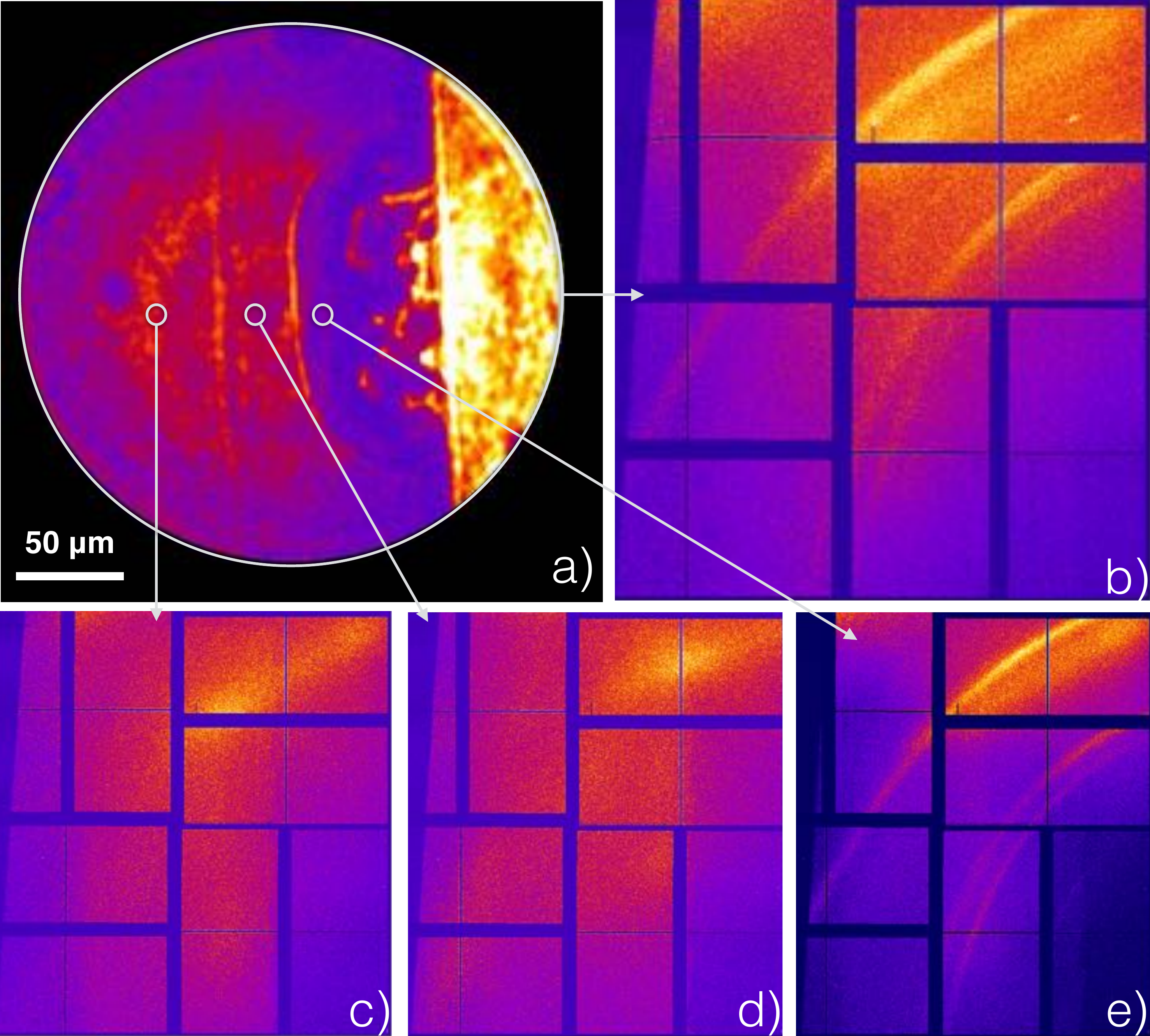} 
 \caption{a) Image of shock wave in single-crystal silicon. Both an elastic wave and a plastic deformation with clear increase in density and  phase transition to polycrystalline Si can be seen. x-ray diffraction when different parts of the image are illuminated are shown by the white circles. b) x-ray diffraction data taken simultaneous with the PCI image in a). The whole image is illuminated with x-rays. c) x-ray diffraction of the cold single crystal Si before the shock wave, taken on a subsequent shot. d) x-ray diffraction of the Si after the elastic shock wave. e) x-ray diffraction of the Si after the plastic wave. The selective illumination that is possible with the PCI instrument allows for determination of structure in different places in the shocked image.}
 \label{fig:pci2}
 \end{figure}

During commissioning of the instrument, phase-contrast images were taken of shock waves traveling through silicon samples. The shocks were driven through the silicon with the MEC glass laser in the orthogonal geometry described in section~\ref{sec:pci} (see Fig.~\ref{fig:pci1}).  The results are shown in Fig.~\ref{fig:pci2}.a). The laser hits the silicon sample on the right of the sample, with a  relative timing with respect to the x-rays that can be electronically changed. An elastic shock front followed by a plastic front with corresponding phase change can be easily distinguished. In principle, any phenomenon that has sufficient contrast, either in phase or intensity can be imaged, and experiments that look at e.g. void collapses, phase transformations and spallation have been proposed and/or performed. Care needs to be taken that the samples are transparent enough: sufficient photons will need to be detectable on the x-ray cameras to determine the contrast with enough dynamic range. Therefore, a judicious choice needs to be made each experiment regarding sample thickness, photon energy, x-ray camera, field of view and resolution, such that sufficiently high quality data can be obtained. One also needs to take into account that only a one-dimensional projection is recorded, and any phase or intensity changes \emph{along} the x-ray beam are not recorded; care needs to be taken that either these changes are not important to the experiment that is being undertaking, or that symmetry considerations can be taken into account to relieve the ambiguity. 

The X-ray beam that is used to image the sample of course also diffracts. We can easily capture this large angle diffraction pattern on X-ray detectors. In this way, we obtain information on the atomic structure of the material that is present in the PCI image. Such diffraction pattern, recorded on a CSPAD detector positioned close to the sample is shown in Fig.~\ref{fig:pci2}.b). It should be noted that this pattern combines the diffraction of the different regions (i.e., both before and after the different  shock fronts) that are visible in the whole Field of View (FOV) of the PCI image in Fig.~\ref{fig:pci2}.a), and is recorded simultaneously  (i.e., with the same x-ray pulse). We have however the option to reduce the FOV by moving the optics table of the instrument (which reduces the value of $\Delta x$ in Fig.~\ref{fig:belenses}), and limit the area from which diffraction occurs. While the PCI image on these subsequent shots may not be very interesting, we can use this to identify the crystallographic phase behind the different shock fronts by analysing the diffraction pattern. This is shown in Figs.~\ref{fig:pci2}.c) -- e), where we reduced the FOV, and therefore the region where diffraction occurs, to approximately 5 $\mu$m, and probed the regions indicated in Fig.~\ref{fig:pci2}.a) by the small white circles, by appropriately aligning the sample, in three subsequent X-ray shots on fresh targets.
In Fig.~\ref{fig:pci2}.c) we probed before the  elastic wave, and the image is indistinguishable from diffraction from an un-driven target (not shown), and no Debye-Scherrer rings are seen.
Probing behind the first shock front ( Fig.~\ref{fig:pci2}.d))  does not reveals rings either, confirming the sample is still a single crystal undergoing elastic compression. Finally, when probing behing the plastic wave (shown in Fig.~\ref{fig:pci2}.e)) reveals Debye-Scherrer rings corresponding to a high pressure phase of polycrystalline silicon. 

The diffraction data allow us to identify the phase transition that occurs behind the imaged shock waves and in this way we obtain information on the sample both on a length scale of a few hundreds of nanometers and the atomic scale.  Although this is beyond the scope of this paper, it is clear that with proper phase retrieval, the change in the transmission in the PCI image can be used to determine the density, spatially resolved and corroborated by the x-ray diffraction, while multiple images taken at different relative time delays can be used to determine the velocity of the poly-crystalline shock fronts, which in principle allows a determination of the equation of state.

\section{Conclusions \label{sec:conclusions}}

The Phase-Contrast Imaging instrument at the Matter in Extreme Conditions endstation of the Linac Coherent Lights Source can image phenomena with spatial resolution of hundreds of nanometers, (currently limited due to the bandwith of the  SASE-beam) and temporal resolution better than 100 femtoseconds. It has the capability to perform ptychographic determination of the x-ray illumination that is used in the phase-contrast imaging experiments. The imaging can be combined with x-ray diffraction for  a simultaneous determination of the atomic structure of the imaged samples and phenomena. The instrument is available for the MEC user community.

\section{Acknowledgements}
We would like to thank Andy Higginbotham for helpfull suggestions.
Parts of this work were funded by Volkswagen Foundation, the DFG under grant SCHR 1137/1-1, and by the German Ministry of Education and Research (BMBF) under grant number 05K13OD2. Use of the Linac Coherent Light Source (LCLS), SLAC National Accelerator Laboratory, is supported by the U.S. Department of Energy, Office of Science, Office of Basic Energy Sciences under Contract No. DE-AC02-76SF00515. The MEC instrument is supported by the U.S. Department of Energy, Office of Science, Office of Fusion Energy Sciences under contract No. SF00515.

\bibliographystyle{apsrev4-1}
\bibliography{pci}
\end{document}